\documentclass[prb,aps,twocolumn,amsmath,amssymb,floatfix,showpacs,
superscriptaddress]{revtex4}

\usepackage[dvips]{graphics}
\usepackage{color}
\definecolor{dred}{rgb}{0,0,0.7}

\begin{document}

\title{\textcolor{dred}{Externally controlled local magnetic field in a 
conducting mesoscopic ring coupled to a quantum wire}}

\author{Santanu K. Maiti}

\email{santanu.maiti@isical.ac.in}

\affiliation{Physics and Applied Mathematics Unit, Indian Statistical
Institute, 203 Barrackpore Trunk Road, Kolkata-700 108, India}

\begin{abstract}

In the present work the possibility of regulating local magnetic field 
in a quantum ring is investigated theoretically. The ring is coupled to
a quantum wire and subjected to an in-plane electric field. Under a finite
bias voltage across the wire a net circulating current is established in
the ring which produces a strong magnetic field at its centre. This magnetic
field can be tuned externally in a wide range by regulating the in-plane 
electric field, and thus, our present system can be utilized to control 
magnetic field at a specific region. The feasibility of this quantum system 
in designing spin-based quantum devices is also analyzed.

\end{abstract}


\maketitle

\section{Introduction}

The phenomenon of voltage driven circular currents in conducting junctions
with single or multiple loop substructures is a notable quantum effect in
low-dimensional systems. Most of the studies involving electron transport
through different bridge systems essentially focus on net current 
transfer~\cite{orella1,orella2,tagami,ventra2,ventra1,aviram,nitzan1,
skm1,skm2,walc,woi,skm3,skm4,skm5,skm6} rather than analyzing current 
distribution among different branches~\cite{dist2,dist3,dist1} of the 
materials within the junction. In last few years, some interesting works
have been done considering different molecular structures where distribution 
of currents in different arms has been analyzed providing the possibilities 
of getting voltage induced circular currents~\cite{cir5,cir4,cir1,cir2,cir3}. 
These circular currents produce substantial magnetic fields at ring centre 
and can be exploited in many ways to explain several interesting quantum 
mechanical phenomena as well as to design quantum devices for future 
applications. 

Controlling of a single spin placed at or near the centre 
of a ring-shaped geometry by means of magnetic field associated with 
circular currents may be the most suitable application towards this 
direction, since proper spin control is extremely important in designing 
spin-based quantum devices~\cite{dev1,dev2,dev3,dev4}. Nowadays people are 
highly focused in computing with single electron spin since it involves 
much lower power dissipation rather than traditional computing which is 
always charge based. In traditional computing, computable informations are 
encoded by electrical charge which has only a magnitude but no direction 
i.e., a scalar quantity. Thus, to encode binary logic bits $0$ and $1$ 
using electrical charge two different amounts of charge are required. 
Now, if the bit is switched the magnitude of the charge required to encode
logic levels should be changed, and accordingly, a net current flow takes 
place. This net current flow certainly produces a power loss. On the other 
hand, if computable informations are encoded by using {\em spin}, which is 
a pseudovector, then the bits $0$ and $1$ can be described by up and down 
spin configurations, respectively. In this case the switching between two 
bits can be associated with the flipping of the spin without transferring 
any net charge and thus much lower power dissipation can be achieved. 

In order to design spin-based quantum devices proper control of magnetic 
field at a particular point is highly important. Till now very few attempts 
have been made to get controlled magnetic field in some particular region 
considering some quantum systems for possible spintronics applications.
\begin{figure}[ht]
{\centering \resizebox*{7.5cm}{5cm}{\includegraphics{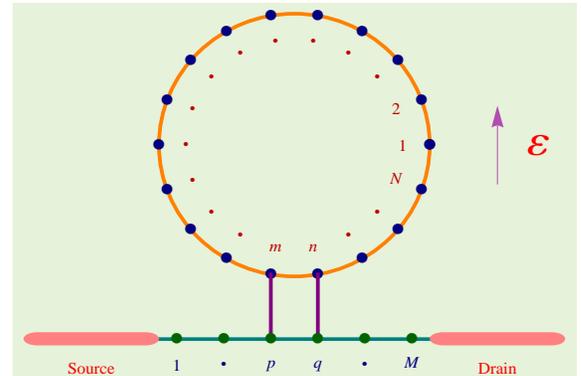}}\par}
\caption{(Color online). A quantum wire coupled to a quantum ring is
attached to source and drain electrodes. An in-plane electric field
$\mathcal{E}$, perpendicular to the quantum wire, is applied to the ring.}
\label{model}
\end{figure}
For example, in 2003 Cho {\em et al.}~\cite{cho} have proposed a double 
quantum dot (QD) system which can act as a magnetically polarized device 
when coherent electrons circulate through closed path of the dots and side 
attached electrodes. It has been shown that manipulating the energy level 
position of each dot, the double QD device can be magnetized as up-, down- 
or non-polarized. In presence of a finite bias, this system exhibits a 
circular current and reaches a maximum of $\sim 0.35$ nA. In the subsequent 
year Lidar {\em et al.}~\cite{lidar} have established a new method to 
produce exponentially
decaying magnetic field using an array of current-carrying wires where
the neighboring wires carry currents in opposite directions. They have
produced a peak magnetic field of $10$ mT. The major issue of this setup 
is the heating effect and it has been examined that the system works
well when the system temperature $T << 2.4$ mK. Such a small temperature 
can be substantiated
with dilution refrigeration technology. In other work, Pershin {\em et al.} 
have shown how to create magnetic field at a specific region of an isolated 
semi-conducting quantum ring using phase-locked infrared laser 
pulses~\cite{pershin}. Here, the induced magnetic field is controlled by 
laser pulses and the estimated magnetic field at the ring center is
of the order of $3$ mT, comparable to the result obtained in 
Ref.~\cite{lidar}. Later, in 2012 Anda {\em et al.}~\cite{anda} have 
presented a system constituted by two QDs embedded in a metallic ring
coupled to external electrodes in which the possibility of controlling
circulating currents by means of external gate potential has been analyzed.
This system could be utilized to regulate circular current induced magnetic
field, but this aspect has not been addressed anywhere in their work. 
Therefore, it can be usually asked that how much magnetic field will be 
established at the center of this ring and how it varies with external
gate potential? Also the feasibility of utilizing this system to design 
a spin-based quantum device has not been discussed.

Though few works are available in this particular area, still several 
important issues are unexplored. For example (i) designing of a simple
quantum system that can produce strong enough circular current, (ii) 
generation of a strong magnetic field at a particular region, and (iii)
control of this magnetic field in a wide variable range. Simultaneously, 
the usefulness of the system even at moderate temperature is also an 
important issue. To the best of our knowledge these features are 
unaddressed so far, and, in the
present work we essentially focus towards this direction. Here, we describe 
a new technique to control magnetic field {\em externally} in a conducting 
quantum ring (QR) which is coupled to a quantum wire (QW) and
subjected to an in-plane electric field. Under a finite bias voltage
a net circulating current is established in the ring and it produces a
strong magnetic field at its centre. This magnetic field can be regulated by 
tuning the in-plane electric field without directly changing other physical 
parameter of the model considered. From our analysis we will show that
the induced magnetic field can be varied in a wide range starting from 
zero to several milliTesla with the help of external in-plane electric 
field.

The phenomenon of circulating current in a conducting ring has already been
established few decades ago in another context~\cite{pc1,pc2,pc3,pc4,
pc5,pc6,pc66,pc7,pc8} where current is induced by Aharonov-Bohm (AB) flux 
$\phi$. This circularly current, the so-called persistent current, in AB 
ring is quite different in origin compared to the circulating currents in 
conducting junctions where current is driven by applied voltage bias. 
In presence of a finite bias, a net circulating current can be obtained 
in a loop geometry even in the absence of any AB flux $\phi$.

We organize the paper as follows. In Section II we describe the model 
together with theoretical formulations for the calculations. Essential 
results are described in Section III, which include the variation of
internal current distribution, effects of system sizes and other physical
parameters on circular current and associated magnetic field, and the
feasibility condition for practical realizations. Finally, we summarize 
our findings in Section IV.

\section{Model and theoretical framework}

\subsection{Tight-binding model}

Our system comprises a quantum ring which is coupled to a quantum wire and
subjected to an in-plane electric field $\mathcal{E}$ perpendicular to the 
wire. The wire is attached to two semi-infinite one-dimensional metal 
electrodes, usually known as source and drain. The schematic diagram of our 
system is presented in Fig.~\ref{model}. In order to get a non-vanishing 
circular current in the ring, we couple it to the wire through two vertical 
bonds. No net current will be obtained in the ring if it is connected to the
wire via single vertical bond due to mutual cancellation of currents moving 
clockwise and anti-clockwise directions. 

We use a tight-binding framework 
to describe the model which is extremely suitable for describing electron 
transport through a conducting junction especially for the case where 
electron-electron interaction is not taken into account.
The single particle tight-binding Hamiltonian that involves the ring,
wire and side-attached electrodes can be written as,
\begin{equation}
H=H_{\mbox{\tiny c}} + H_{\mbox{\tiny el}} + H_{\mbox{\tiny tn}}.
\label{equ1}
\end{equation}
where, $H_{\mbox{\tiny c}}$, $H_{\mbox{\tiny el}}$ and $H_{\mbox{\tiny tn}}$
correspond to different sub-Hamiltonians those are described as follows. The 
first term $H_{\mbox{\tiny c}}$ corresponds to the Hamiltonian of the 
conductor bridging the two electrodes i.e., the ring with coupled quantum 
wire. Under nearest-neighbor hopping approximation the Hamiltonian 
$H_{\mbox{\tiny c}}$ gets the form,
\begin{eqnarray}
H_{\mbox{\tiny c}} & = & \sum_i \epsilon_i^r c_i^{r\dagger} c_i^r + 
\sum_i t_r \left[c_{i+1}^{r\dagger} c_i^r + c_i^{r\dagger} c_{i+1}^r\right] 
\nonumber \\
 & + & \sum_i \epsilon_i^w c_i^{w\dagger} c_i^w + \sum_i t_w
\left[c_{i+1}^{w\dagger} c_i^w + c_i^{w\dagger} c_{i+1}^w\right] \nonumber \\
 & + & \lambda \left[c_m^{r\dagger}c_p^w + c_p^{w\dagger}c_m^r
       + c_n^{r\dagger}c_q^w + c_q^{w\dagger}c_n^r\right]
\label{equ2}
\end{eqnarray}
where, $c_i^{r\dagger}$ and $c_i^r$ represent the creation and annihilation
operators of an electron at $i$-th site of the QR, while for the QW these 
operators are represented by $c_i^{w\dagger}$ and $c_i^w$, respectively. 
The ring is parametrized by on-site energy $\epsilon_i^r$ and nearest-neighbor
hopping integral $t_r$, and similarly we parametrize the wire by the terms
$\epsilon_i^w$ and $t_w$. $\lambda$ measures the coupling between the QR and 
QW, where they are coupled through the lattice sites $m$, $n$, $p$ and $q$ 
(see Fig.~\ref{model}), those are variable.

Now, in presence of an in-plane electric field which is perpendicular to the
wire, the site energy of the QR becomes field dependent. For a $N$-site
ring it becomes: $\epsilon_i^r=(e a N\mathcal{E}/2\pi)\cos[2\pi(i-1)/N]$, 
where $e$ corresponds to the electronic charge, $a$ represents the lattice 
spacing and $\mathcal{E}$ measures the strength of the in-plane electric 
field. This field dependent site energy relation is again simplified by
introducing the dimensionless electric field strength $\xi$ and can be 
written as $\epsilon_i^r=(\xi t_w N/2\pi) \cos[2\pi(i-1)/N]$, where
$\xi=e a \mathcal{E} /t_w$. In our formulation, we use the parameter $M$
to describe the system size of the QW.

The second term in Eq.~\ref{equ1} corresponds to the Hamiltonian of 
one-dimensional source and drain electrodes. It is expressed as
\begin{eqnarray}
H_{\mbox{\tiny el}} & = & H_{\mbox{\tiny S}} + H_{\mbox{\tiny D}} 
\nonumber \\
 & = & \sum_{\alpha={\mbox{\tiny S,D}}} \left\{\sum_n \epsilon_0 
d_n^{\dagger} d_n 
+ \sum_n t_0 \left[d_{n+1}^{\dagger} d_n + h.c. \right]\right\}, 
\nonumber \\
\label{equ3}
\end{eqnarray}
where, $\epsilon_0$ presents the on-site energy and $t_0$ gives the
nearest-neighbor hopping strength in the electrodes. For a $n$-th site
electron in these electrodes the creation and annihilation operators are 
described by $d_n^{\dagger}$ and $d_n$, respectively.

Finally, the third term in Eq.~\ref{equ1} denotes the coupling of the wire 
to the source and drain electrodes. Considering $\tau_{\mbox{\tiny S}}$ and 
$\tau_{\mbox{\tiny D}}$ are coupling integrals between the wire and 
side-attached electrodes, the Hamiltonian $H_{\mbox{\tiny tn}}$ can be 
written as,
\begin{eqnarray}
H_{\mbox{\tiny tn}} & = & H_{\mbox{\tiny S,wire}} + H_{\mbox{\tiny D,wire}} 
\nonumber \\
& = & \tau_{\mbox{\tiny S}}[c_1^{w\dagger}d_0 + h.c.] + \tau_{\mbox{\tiny D}}
[c_M^{w\dagger}d_{M+1} + h.c.].
\label{equ4}
\end{eqnarray}

\subsection{Circular current in the QR}

\begin{figure}[ht]
{\centering \resizebox*{6cm}{5.2cm}{\includegraphics{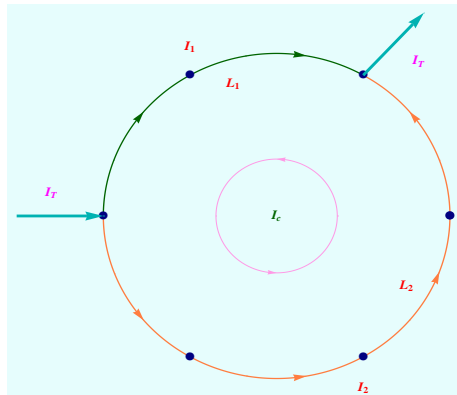}}\par}
\caption{(Color online). Schematic view of current distribution in a ring 
geometry coupled to two electrodes. The filled blue circles correspond to 
the positions of the atomic sites. A net circular current $I_c$ is 
established in the ring.}
\label{circular}
\end{figure}

To determine circular current in the system considered let us first focus
on the current distribution in a simple model where a mesoscopic ring is 
coupled to two electrodes (see Fig.~\ref{circular}). A net junction current 
$I_T$ flows between source and drain, where $I_1$ and $I_2$ are the currents 
propagating through upper and lower arms of the ring, respectively. For
the current flowing in the counter-clockwise direction we use positive 
sign, while it is negative for the other direction. Following the current
distribution given in Fig.~\ref{circular}, the net circulating current in
the ring is defined as~\cite{cir1},
\begin{equation}
I_c=\frac{1}{L} (I_1 L_1 + I_2 L_2)
\label{equ5}
\end{equation}
where, $L=L_1+L_2$. $L_1$ and $L_2$ are the arm lengths. This is the basic
definition of circular current in any loop geometry attached to two 
electrodes. The relation, Eq.~\ref{equ5}, immediately suggests that for a
symmetrically connected ring no net circular current will appear since in
this case $L_1=L_2$ and $I_1=-I_2$. Now to calculate $I_c$ using 
Eq.~\ref{equ5}, we have to determine currents in different segments of the
ring geometry. We compute these currents using Green's function formalism.
At absolute zero temperature (T=$0\,$K), current in a bond connecting the
sites $i$ and $j$, where $j=i\pm 1$, can be expressed as,
\begin{equation}
I_{ij} = \displaystyle \int \limits_{E_F-\frac{eV}{2}}^{E_F+\frac{eV}{2}}
J_{ij}(E) \, dE.
\label{equ6}
\end{equation}
$J_{ij}(E)$ is the charge current density and it provides a net bond current
upon integrating over a particular energy window. $E_F$ is the equilibrium
Fermi energy and $V$ is the applied bias voltage between source and drain 
electrodes. To describe current density we impose correlation Green's 
function $G^n$ and in terms of it, $J_{ij}(E)$ can be written as~\cite{dist3}
\begin{equation}
J_{ij}(E) = \frac{4e}{h} \mbox{Im} \left[(H_{\mbox{\tiny c}})_{ij}G_{ij}^n 
\right].
\label{equ7}
\end{equation}
The correlation Green's function $G^n$ is defined in the form: 
$G^n=G^r \Gamma_{\mbox{\tiny S}} G^a$, where $G^r$ and $G^a$ are the 
retarded and advanced Green's functions, respectively, and they obey the
relation $G^r=(G^a)^{\dagger}$. $\Gamma_{\mbox{\tiny S}}$ is the coupling
matrix associated with the coupling of the conductor to the source electrode. 
Introducing the concept of contact self-energies $\Sigma_{\mbox{\tiny S}}$ 
and $\Sigma_{\mbox{\tiny D}}$ due to two semi-infinite one-dimensional 
electrodes we can write~\cite{datta},
\begin{equation}
G^r=\left(E-H_{\mbox{\tiny c}}-\Sigma_{\mbox{\tiny S}} - 
\Sigma_{\mbox{\tiny D}} \right)^{-1}
\label{equ8}
\end{equation} 
where, $E$ is the energy of an injecting electron. During evaluation of
the correlation function $G^n$ we fix the occupation function of the source
electrode to unity while for the drain electrode it becomes zero.

\subsection{Magnetic field in the QR associated with circular current}

Due to circular current $I_c$, a net magnetic field is established in the
ring. At any point $\vec{r}$ inside the ring, we calculate the magnetic
field by using Biot-Savart's law which looks like~\cite{cir1},
\begin{equation}
\vec{B}(\vec{r})=\sum_{(i,j)} \int \frac{\mu_0}{4\pi} I_{ij} 
\frac{d\vec{r} \times (\vec{r}-\vec{r^{\prime}})}
{|(\vec{r}-\vec{r^{\prime}})|^3},
\label{equ9}
\end{equation}
where, $\vec{r^{\prime}}$ is the position vector for the bond current 
element $I_{ij}d\vec{r^{\prime}}$ and $\mu_0$ is the magnetic constant.

Throughout the results discussed below in Sec. III, unless otherwise stated, 
we fix the electronic temperature to zero and assume that the entire voltage 
is dropped at the wire-to-electrode interfaces. The other common parameters 
are as follows: $\epsilon_i^r=\epsilon_i^w=0$ $\forall$ $i$, 
$t_r=t_w=1$ eV, $\tau_{\mbox{\tiny S}} = \tau_{\mbox{\tiny D}} =1$ eV, 
$\epsilon_0=0$ and $t_0=2$ eV. The lattice spacing $a=1\, A^{\circ}$ and the
equilibrium Fermi energy $E_F$ is set at zero.

\section{Numerical results and discussion}

Based on the above theoretical framework we are now ready to present
our numerical results for circular currents and associated magnetic fields,
and the effect of in-plane electric field on them. These results are 
described in different sub-sections as follows.

\subsection{Internal current distribution}

\begin{figure}[ht]
{\centering \resizebox*{8cm}{5.5cm}{\includegraphics{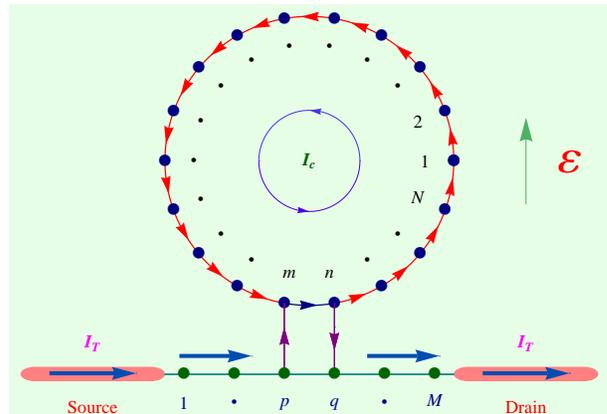}}\par}
\caption{(Color online). Current distribution in different segments of the
bridge setup at the bias voltage $V=1$ V when the dimensionless electric 
field strength is set equal to $\xi=0.5$. Here we consider $N=18$, $M=8$, 
$m=14$, $n=15$, $p=4$, $q=5$ and $\lambda=1$ eV. The arrows correspond 
to the propagation directions of the currents in different branches, while 
their magnitudes are described by different colors of the arrows. A net 
circulating current $I_c$ is established in the ring.}
\label{dist}
\end{figure}

Before focusing to the central problem i.e., the possibility of controlling 
local magnetic field associated with voltage induced circular current by 
means of external electric field, without directly changing other system 
parameters, we start with the current distribution shown in Fig.~\ref{dist}. 
A net junction current $I_T$ flowing from the source to drain electrode is 
distributed among different branches of the bridge setup where distinct 
colored arrows correspond to different magnitudes of the bond currents. 
These bond currents are determined when the bias voltage is fixed at $V=1$ V 
and the dimensionless electric field strength is set equal to $\xi=0.5$.
In this setup we consider a $18$-site ring so that the upper arm contains 
$17$ identical bonds, while only a single bond exists in the lower arm, and
the currents in these two arms with unequal magnitudes associated with the 
voltage bias are propagating in a particular direction (anti-clockwise) 
which result a non-vanishing circular current $I_c$ in the ring. For a 
finite bias, the direction of these bond currents strongly influenced by 
the way the QR is coupled to the QW. In our chosen configuration, illustrated
in Fig.~\ref{dist}, two neighboring atomic sites of the ring are coupled
to the wire such that the difference ($L_1-L_2$) between the arm lengths
becomes maximum. With reducing this length difference, oppositely rotating 
bond currents in the upper and lower arms of the QR can be obtained depending 
on the voltage window, and eventually when $L_1-L_2=0$ currents with exactly
equal in magnitude are available in the upper and lower arms of the ring
those propagate in reverse directions irrespective of the choice of the
voltage bias which provide a vanishing circular current in the QR. The 
phenomenon of voltage driven circular current in a conducting loop can be
\begin{figure}[ht]
{\centering \resizebox*{8cm}{5.5cm}{\includegraphics{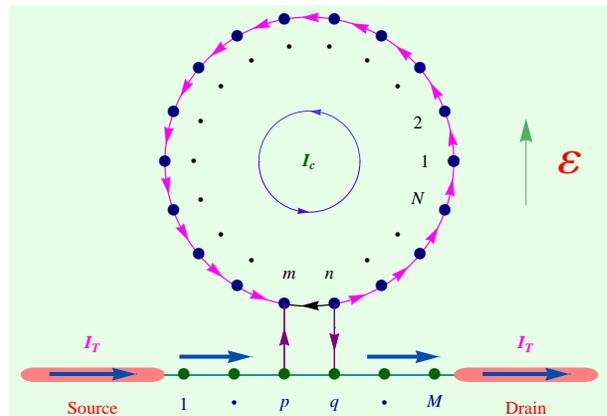}}\par}
\caption{(Color online). Current distribution in different segments (shown
by different colored arrows) when the dimensionless electric field strength
is fixed at $\xi=0.25$ and the ring-to-wire coupling is fixed at 
$\lambda=1.25$ eV. All the other physical parameters remain same as taken
in Fig.~\ref{dist}.}
\label{dist1}
\end{figure}
explained in terms of current carrying states associated with energy 
eigenvalues of the system. For a finite voltage bias when a single
resonating state lies within the voltage window circulating current 
is obtained due to this particular state and the direction of this current 
is also governed by it. On the other hand, when more resonating states 
those carry currents in opposite directions appear within the voltage 
regime, then all of them contribute to the current and a resultant is 
obtained. Certainly, the sign of the net current depends on the dominating 
states i.e., the states contributing more to the current than the others.

Like the voltage bias and the ring-to-wire configuration ($L_1 \sim L_2$),
nature of these bond currents is also affected noticeably by the other 
physical parameters of the system. To substantiate it, in Fig.~\ref{dist1} 
we present the variation of current distribution in individual branches
of the same bridge configuration as taken in Fig.~\ref{dist}, except that
$\xi=0.25$ and $\lambda=1.25$ eV. Under this situation oppositely rotating
currents with unequal magnitudes are available even when $L_1-L_2$ is 
maximum, and, depending on the branch currents a net circulating current 
is established in the QR.

\subsection{Circular current and associated magnetic field for different 
system sizes}

Following the above features of current distribution among different 
branches we can now move to demonstrate the characteristic features 
of circular current and associated magnetic field as a function of
\begin{figure}[ht]
{\centering \resizebox*{8cm}{5cm}{\includegraphics{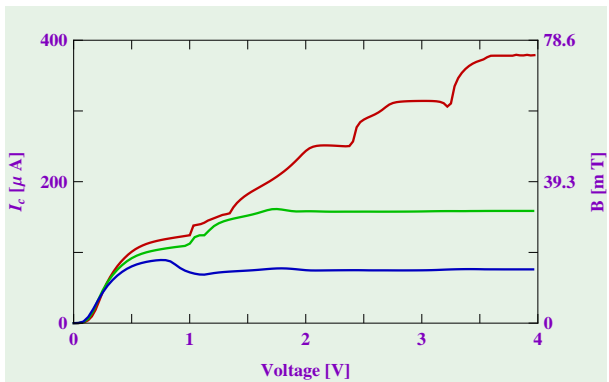}}\par}
\caption{(Color online). Circular current ($I_c$, left axis) and the 
associated magnetic field $B=(0,0,B_z)$ (right axis) at the ring centre 
as a function of applied bias voltage for different values of dimensionless 
electric field strength $\xi$. Here we set $N=20$, $M=10$, $m=15$, $n=16$, 
$p=5$, $q=6$ and $\lambda=1$ eV. The red, green and blue lines correspond 
to $\xi=0$, $0.42$ and $0.5$, respectively.}
\label{currN20}
\end{figure}
voltage bias, and, the effect of in-plane electric field on these quantities
for different system sizes.

As representative example, in Fig.~\ref{currN20} we present the variation of
circular current $I_c$ (left axis) and the associated magnetic field 
$B=(0,0,B_z)$ (right axis) at the ring centre considering $N=20$ and $M=10$.
The results are computed for three distinct values of the dimensionless 
electric field strength $\xi$, those are illustrated by three different 
colored curves. Several interesting features are observed. First, in the
absence of in-plane electric field a very large circulating current
(red line) is obtained in the QR and it gradually increases with the bias
voltage and reaching a maximum of $\sim 400$ $\mu A$. The current also
exhibits some tiny oscillations with the applied bias. Second, the associated
magnetic field $B$ at the ring center induced by the circular current is 
considerably large and reaches to $\sim 77$ mT. This is reasonably high 
compared to the previous studies of externally controlled magnetic field
at a particular region where local magnetic field is generated either by 
using laser-controlled approach~\cite{pershin} or with the help of arrays 
of parallel current-carrying wires~\cite{lidar} or using double QD 
device~\cite{cho}. Finally, both the circular 
current and induced magnetic field decrease with increasing dimensionless 
electric field strength $\xi$, and, for a wide voltage regime they are 
almost constant.

The above three features can be physically understood as follows. As 
discussed in the previous sub-section we note that the circular current 
$I_c$ for a particular bias is a resultant contribution of all the states
carrying currents those come into the voltage window. Therefore, with 
increasing external bias more and more resonating states appear and all
of them contribute to the current which finally yield a larger circular
current. Obviously, a strong magnetic field is obtained at the ring center
\begin{figure}[ht]
{\centering \resizebox*{8cm}{5cm}{\includegraphics{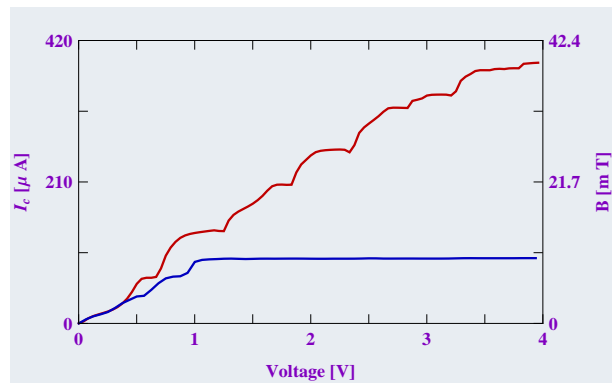}}\par}
\caption{(Color online). Voltage dependence of circular current ($I_c$, left
axis) and the associated magnetic field $B=(0,0,B_z)$ (right axis) at the
ring centre for different values of dimensionless electric field strength
$\xi$. Here we set $N=40$, $M=20$, $m=30$, $n=31$, $p=10$, $q=11$ and 
$\lambda=1$ eV. The red and blue lines correspond to $\xi=0$ and $0.25$, 
respectively.}
\label{currN40}
\end{figure}
associated with this large current $I_c$. The reduction of the circular 
current as well as the induced magnetic field in presence of in-plane 
electric field can be illustrated by considering the QR as a disordered
one. In the absence of any in-plane electric field site energies of the
QR are equal ($\epsilon_i^r=0$ $\forall$ $i$) which make the ring as a 
perfect one. Whereas, for finite electric field ($\mathcal{E} \ne 0$) 
these site energies are field dependent and they are no longer identical 
with each other which provide a disordered QR. Under this situation the 
current carrying states are not as much extended like the electric field 
free case ($\mathcal{E}=0$) and hence reduced circular current is obtained. 
Accordingly, we also get less magnetic field at the ring center. With
increasing electric field strength both the circular current and magnetic 
field get reduced which can be detected from the curves shown in
Fig.~\ref{currN20}. More clear picture of this field dependence on circular
current and associated magnetic field is discussed in a separate sub-section
(sub-section~D) below. In addition to the above facts it is also interesting
to note that for non-zero electric field the circular current gradually 
increases with external bias in a relatively narrow voltage region, while
the rate of increment sharply decreases in the higher voltage regimes
(see the green and blue lines in Fig.~\ref{currN20}). The reason is that
in the weak voltage regions, resonant states contribute significantly
differently which result a net larger current, while the difference gets
reduced slowly with higher voltage regions associated with the split
\begin{figure}[ht]
{\centering \resizebox*{8cm}{5cm}{\includegraphics{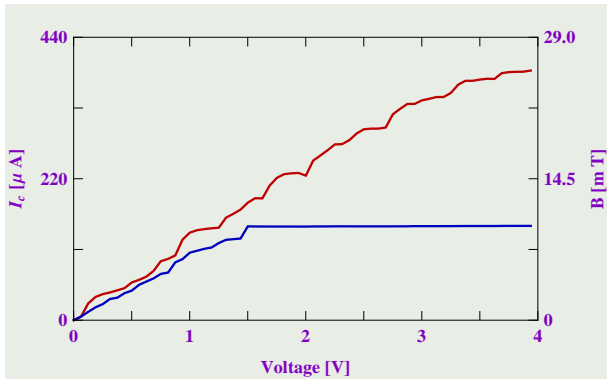}}\par}
\caption{(Color online). Voltage dependence of circular current ($I_c$, left
axis) and the associated magnetic field $B=(0,0,B_z)$ (right axis) at the
ring centre for different values of $\xi$. Here we set 
$N=60$, $M=30$, $m=45$, $n=46$, $p=15$, $q=16$ and $\lambda=1$ eV. The red
and blue lines correspond to $\xi=0$ and $0.15$, respectively.}
\label{currN60}
\end{figure}
energylevels and hence almost constant current is generated. Finally, the 
behavior of tiny oscillations in circular current can be elucidated in 
terms of the discreteness of resonating energy eigenstates for the finite 
size systems. The bond current is obtained by integrating the charge 
current density (see Eq.~\ref{equ6}) over a particular voltage window 
where finite resonant states contribute to the current. In the absence 
of electric field, sharp resonant states are obtained since the QR behaves 
like an impurity-free ring which generate tiny oscillating current in the 
integration procedure, while in the case of finite electric field much 
broader resonant energy levels arise as the QR is treated as a disordered 
one and a less oscillating current appears.

Observations of circular current and associated magnetic field with other
sizes of the QR and QW are qualitatively similar to those with the bridge
setup where we set $N=20$ and $M=10$ (Fig.~\ref{currN20}). The results 
are shown in Figs.~\ref{currN40} and \ref{currN60} for different strengths
of the dimensionless electric field $\xi$. In Fig.~\ref{currN40} we set
$N=40$ and $M=20$, while in Fig.~\ref{currN60} we choose $N=60$ and $M=30$.
Comparing the spectra shown in Figs.~\ref{currN20}-\ref{currN60} we 
emphasize that the results are quite robust and can be utilized to
generate strong circular current as well as magnetic field.

\subsection{Effect of $\lambda$ on circular current and magnetic field}

\begin{figure}[ht]
{\centering \resizebox*{8cm}{5cm}{\includegraphics{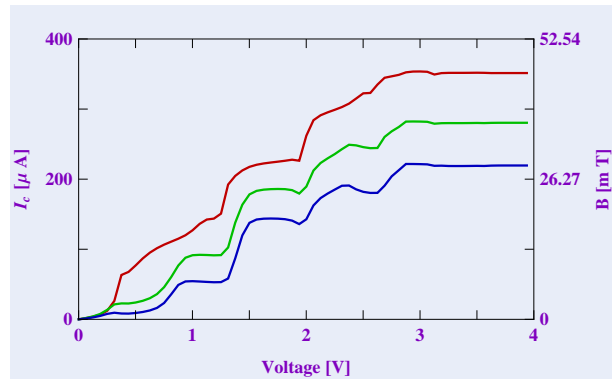}}\par}
\caption{(Color online). Circular current ($I_c$, left axis) and the 
associated magnetic field $B=(0,0,B_z)$ (right axis) at the ring centre as
a function of external bias for different ring-to-wire coupling strengths 
($\lambda$) when
the dimensionless electric field strength is fixed at $\xi=0.1$. The other 
physical parameters are: $N=30$, $M=12$, $m=23$, $n=24$, $p=6$ and $q=7$. 
The red, green and blue lines correspond to $\lambda=0.8$ eV, $1.2$ eV and 
$1.3$ eV, respectively.}
\label{lambdaN30}
\end{figure}

In order to substantiate the effect of ring-to-wire coupling ($\lambda$) 
on circular current and associated magnetic field induced by this current, 
in Fig.~\ref{lambdaN30} we present the results for a $30$-site ring 
considering different values of $\lambda$. The results are computed setting
the dimensionless electric field strength $\xi=0.1$, where the red, green 
and blue lines correspond to $\lambda=0.8$ eV, $1.2$ eV and $1.3$ eV, 
respectively.
From the spectra it is observed that the circular current and its associated
magnetic field get slightly reduced with increasing the ring-to-wire
coupling strength, though all the other features remain qualitatively 
similar as discussed earlier.

\subsection{Circular current and magnetic field versus electric field
characteristics}

\begin{figure}[ht]
{\centering \resizebox*{8cm}{5cm}{\includegraphics{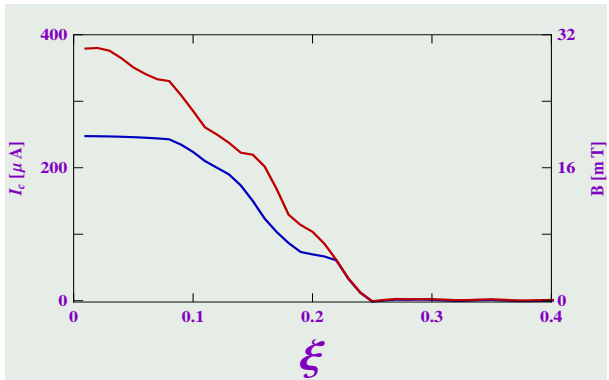}}\par}
\caption{(Color online). Circular current ($I_c$, left axis) and the 
induced magnetic field $B=(0,0,B_z)$ (right axis) at the ring center
as a function of dimensionless electric field strength $\xi$ for different 
values of bias voltage. Here we take $N=50$, $M=20$, $m=38$, $n=39$, $p=10$,
$q=11$ and $\lambda=1$ eV. The red and blue curves correspond to $V=3.5$ V 
and $2$ V, respectively.}
\label{fld}
\end{figure}

The results presented so far to regulate circular current induced magnetic
field at the ring center by means of external in-plane electric field are
computed for some typical field strengths. To have a fine tuning of local
magnetic field at ring center with the help of external electric field, now
we focus our attention on the results shown in Fig.~\ref{fld}. Here we
demonstrate the behavior of circular current (left axis) and induced magnetic 
field (right axis) as a function of dimensionless electric field strength
$\xi$ considering $N=50$ and $M=20$. The results are shown for two different
bias voltages, where the red and blue curves correspond to $V=3.5$ V and
$2$ V, respectively. Quite interestingly we find that both circular
current and the corresponding magnetic field decrease monotonically with 
increasing the strength of the electric field. This reduction can be 
easily justified from our previous discussion where the QR is treated
as a disordered one in presence of the electric field. Certainly, the 
ring becomes more disordered with increasing the field strength which 
results more reduced current, and eventually, it drops to zero for strong
field strength.
Our results clearly suggest that the magnetic field at the ring centre 
can be well adjusted by tuning the external electric field, and thus, our 
proposed quantum system can be used as an externally controlled source 
of local magnetic field.

\subsection{Practicability consideration and technological importance}

Finally, we discuss the applicability of this quantum system to
develop spin-based quantum devices. To get it feasible, first it is
important to calculate the required magnetic field $B$ for a single-qubit
operation. It is well known that magnetic field can able to change the
alignment of a single spin. To rotate a single spin by a relative angle
$\theta$ for a time scale $\tau$ the required magnetic field is given
by~\cite{lidar}: $B=2 \hbar \theta/g \mu_B \tau$, where $g$ is the 
$g$-factor and $\mu_B$ is the Bohr magneton. Assuming the average operation 
time~\cite{lidar} $\tau=5$ ns, the desired magnetic field for rotating 
the spin by an angle $\theta=\pi/2$ is $B \sim 7$ mT.

For the present system we find that the induced magnetic field at the 
ring centre
is quite high compared to the above estimated magnetic field to rotate a
single spin by an angle $\theta=\pi/2$. In addition to this, we also 
establish that the magnetic field induced by circular current can be well
controlled and varied in a wide range starting from zero to several 
milliTesla (mT) by means of external in-plane electric field, without 
directly disturbing other physical parameters of the system. Our results 
clearly suggest that in all respects the present quantum system is 
technologically feasible and can be utilized to design spin-based quantum 
devices.

\section{Conclusion}

To conclude, in the present work, we have demonstrated one possible route
of controlling local magnetic field at a particular point {\em externally}
without disturbing the system parameters. Our system comprises a quantum 
ring which is directly coupled to a quantum wire under a specific 
configuration. The ring is subjected to an in-plane electric field and it is
the key controlling parameter of our present study. In presence of a finite 
bias voltage across the two ends of the quantum wire, a net circulating 
current appears in the ring, and accordingly, a magnetic field is 
established at its centre. Changing the in-plane electric field and keeping 
all other parameters unchanged, the magnetic field at the ring centre can 
be adjusted in a tunable way. We have employed a simple tight-binding
framework to illustrate the quantum system and calculated all 
the results using Green's function formalism. Our presented results 
undoubtedly suggest that the system can be utilized as a source of local 
magnetic field that can be controlled {\em externally}. Finally, we have
also discussed about the feasibility of our quantum system in designing 
spin-based quantum devices. We have established that the induced magnetic
field at the ring center can be varied in a wide range with the help of
external in-plane electric field which is highly significant in the 
present era of spintronic applications. We strongly believe that the 
present investigation provides a much simpler way of controlling local 
magnetic field in such a broader range than the conventional 
techniques~\cite{lidar,pershin}.

Some valid approximations have been taken into account in this study. The
first one is the zero temperature approximation. Here we have computed the
results considering zero temperature limit, but these results are equally 
valid even at finite temperatures since thermal broadening of the 
energy levels of the bridging system is much weaker than the broadening 
caused by the wire-to-electrode coupling~\cite{datta,br1,br2,br3}. To
have a quantitative estimate we calculate average spacing $\Delta E$
of the energy levels of the bridging system which is $\sim 0.1$ eV,
and it is reasonably high compared to the thermal energy $k_BT=0.0256$ eV
in the room temperature ($25\,^{\circ}{\rm C}$) limit. Thus, our predicted 
results fit reasonably well at moderate temperatures. The 
other approximation is the consideration of non-interacting electron 
picture. In presence of electron-electron interaction~\cite{ee1,ee2,ee3} 
one might expect some interesting patterns, but all the physical phenomena
analyzed here will remain unchanged. Beside these we have also ignored 
the effects of electron dephasing~\cite{dp1,dp2,dp3,dp4}, system 
impurities~\cite{dis1,dis2}, etc. These issues will be addressed elsewhere 
in our future work.

Finally, it is important to note that during the numerical calculations we 
have taken some specific parameter values to compute the results, but all
the physical properties studied here remain absolutely invariant for other 
choices of the parameter values, which essentially suggests the robustness
of our investigation. Our results certainly demand an experimental 
verification in this line.

\end{document}